\documentclass[11pt, oneside]{article}
\usepackage{geometry}
\usepackage{graphicx}
\usepackage{amssymb}
\usepackage{rotating}
\usepackage{hyperref}

\title{\normalsize RD51-NOTE-2015-012\\
~\\
\huge{Prospects in MPGDs development\\for neutron detection}\\
\normalsize Summary of RD-51 Academia-Industry Matching Event\\Second Special Workshop on Neutron Detection with MPGDs}
\author{Gabriele Croci (University of Milano-Bicocca, INFN \& CNR), \\ Fabrizio Murtas (INFN \& CERN), \\ Filippo Resnati (CERN)\\
\\
Organising committee of the Academia-Industry Matching Event\\
A.~Breskin (Weizmann Institute),
A.~Delbart (CEA),\\
S.~Duarte~Pinto (CERN),
I.~Giomataris (CEA),\\
B.~Guerrard (ILL),
R.~Hall-Wilton (ESS),\\
J.~Le~Goff (CERN),
F.~Murtas (INFN \& CERN),\\
A.~Pacheco (CERN),
L.~Ropelewski (CERN),\\
M.~Titov (CEA),
T.~Tsarfati (CERN)}
\date{}							% Activate to display a given date or no date

\begin{document}
\maketitle

\section{Introduction}
The aim of this document is to summarise the discussion and the contributions from the 2nd Academia-Industry Matching Event on Detecting Neutrons with MPGDs~\cite{neutronEvent2} which took place at CERN on the $16^{th}$ and the $17^{th}$ of March 2015.
The first event of this kind~\cite{neutronEvent1}, organised in 2013, was summarised in~\cite{summary1}.
These events provide a platform for discussing the prospects of Micro-Pattern Gaseous Detectors (MPGDs)~\cite{MPGD} for thermal and fast neutron detection, commercial constraints and possible solutions.
The aim is to foster the collaboration between the particle physics community, the neutron detector users, instrument scientists and fabricants.
This document is not meant to be a comprehensive review of the neutron detection with gaseous detectors, instead it is an addendum and a continuation of the previous summary.

Very good position resolution, high particle flux capability, radiation tolerance, low material budget, large surfaces and low energy threshold are the key features which make MPGDs flexible and widespread devices in High Energy Physics experiments.
These features make them interesting solutions also for the next generation neutron scattering instruments and beam monitors.
The development of \emph{non-standard} neutron detectors, possibly based on MPGDs, is important not only because of the $^{3}$He shortage, which forces to find urgently valuable alternatives, but also to extend the capabilities of the actual detectors (e.g.\ very high particle flux capabilities).

MPGDs are gas-based particle detectors that extend the capabilities of the Multi-Wire Proportional Chambers, even them largely used by the High Energy Physics and neutron scattering communities.
In MPGDs the electrodes to shape the electric field in the gas volume are printed with photolithographic techniques on solid substrates.
Due to the presence of \emph{cathode} electrodes in the vicinity of the amplification region, MPGDs features a fast evacuation of ions, and therefore very good capabilities to cope with high particle fluxes.
The electric field is modestly affected by spurious deposits on the electrodes, reducing very much the \emph{classical} ageing.
The readout electrodes, placed on solid and often rigid substrate, allow position resolutions of the order of tens of micrometer.
Since 1988, when the first detector of this kind was invented, the Micro-Strip Gas Chamber (MSGC)~\cite{Oed:1988}, several MPGDs families have been developed.
The two most common are based on MicroMeGas~\cite{Giomataris:1996} and GEM~\cite{Sauli:1997} solutions.

RD51 is the world's largest collaboration for the development and application of Micro-Pattern Gaseous Detectors with the aim of facilitating the development of advanced gas-avalanche detector technologies and associated electronic readout systems, for applications in basic and applied research.
In this context, RD51 organised a series of Academia-Industry matching events, in order to advertise the capabilities of MPGDs, and to bring out the issues and the constraints for possible MPGDs applications outside the fundamental research.

Neutrons are widely used in modern matter physics in a variety of applications.
From the very beginning, thermal neutron diffraction has been used to study crystals and their properties, exploiting the fact that neutrons have a higher penetration in matter than X-rays, of which they constitute a natural complementary probe.
Neutrons can penetrate samples of large thickness (many cm) to reveal details of the crystal structure of the bulk of the specimen.
On the other hand, neutrons are not stopped by the walls of sample containers, thus allowing testing in high-vacuum or high-pressure conditions.
Moreover, thermal neutrons are widely used to test molecular dynamics, in a way similar to Raman or IR measurements, with the additional advantage of not being subject to EM selection rules and being more sensitive to light elements.
Finally, more recent applications of neutrons as a probe for matter physics are neutron radiography and tomography, neutron Spin-Echo and Neutron Resonance Capture Analysis.
Many neutron sources devoted to matter physics are nowadays available throughout the world, both steady-state (i. e. reactor-based) or pulsed (i. e. accelerator driven).
Exhaustive and comprehensive documents are~\cite{Squires, Windsor}.

This document targets both instrument scientists of beam lines at neutron sources and researcher that need to develop a high rate, large area, low cost fast neutron diagnostic devices for all neutron-linked physics applications (plasma physics, medical physics, space physics, etc.).

\section{Thermal neutron detectors}
Thermal neutrons are used as a probe to investigate the matter properties because they are non-damaging and highly penetrating.
Neutron scattering instruments can measure at the same time both energy and momentum transfer.
Unlike the X-rays, the neutrons scatter against the nuclei of the atoms.
It happens that the total neutron cross-sections are \emph{complementary} to the one of X-rays: Neutrons interact also with very light atoms, and mainly with hydrogen.
Neutron elastic scattering brings information about the position of the atoms, while inelastic scattering gives insights of the atoms dynamics.
%SOMETHING MRE ABOUT THE INSTRUMENTS?

Though not the only production methods, fission (in a reactor) and spallation (in a proton accelerator) are the two most common, efficient and effective ways to produce high intensity neutron beams.
The detail discussion of neutron production, thermalisation and transport is beyond the scope of this document.
%Exhaustive and comprehensive documents are CITATIONS.
\begin{figure}
\centering
\includegraphics[width=0.8\linewidth]{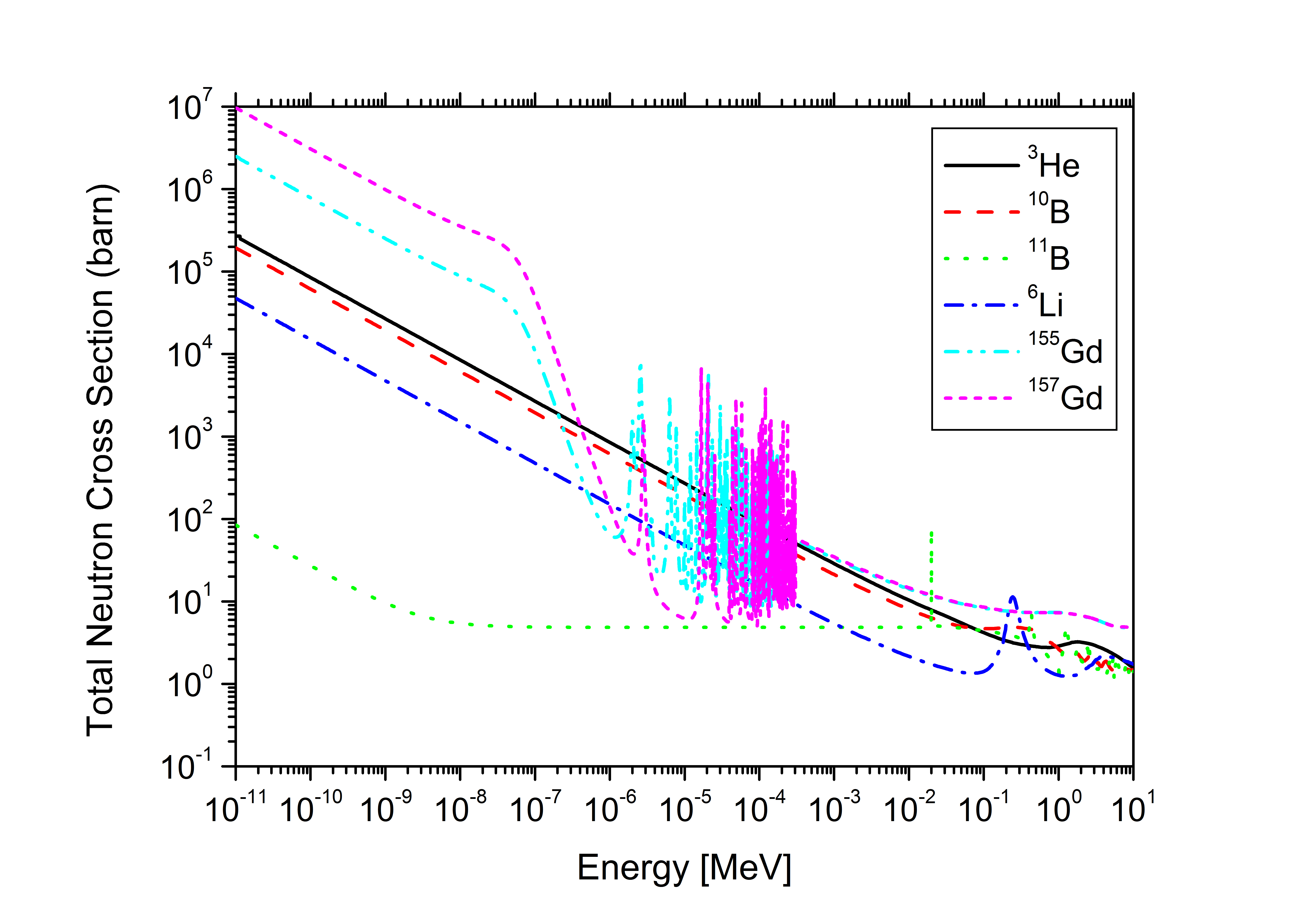}
\caption{Neutron capture cross-sections as a function of the neutron energy for different nuclei.}
\label{fig:CrossSection}
\end{figure}

Thermal neutrons are not energetic enough to be directly detectable via elastic scattering.
The detection principle is based on the absorption and the conversion of the neutron into charged particles via nuclear reactions.
The most exploited, for which the cross-sections as a function of the neutron energy are shown in figure~\ref{fig:CrossSection}, reactions are:
$^{3}$He(n,p)$^{3}$H,
$^{6}$Li(n,$\alpha$)$^{3}$H,
$^{10}$B(n,$\alpha$)$^{7}$Li,
$^{X}$Gd(n,e$^{-}$)$^{X+1}$Gd$^{+}$,
$^{235}$U(n,f).
The converter (in the form of fluid or solid) can either be the active material of the detector, as in the case of $^{3}$He tubes, or simply the target from where the secondary charged particles are emitted.
In MPGDs the typical converter is a set of thin foils usually evaporated or sputtered with the sensitive atoms onto a thin aluminium substrate.
The material, the thickness and the geometry of the converter is optimised for the needed sensitivity to the given energies of the neutrons.

In general within a detector, a neutron may
do not interact at all,
be scattered and potentially be detected far from the first interaction point,
be absorbed, but not detected,
or be properly detected.
This gives rise to a very complex detector behaviour that can usually be summarised with quantities such as
efficiency,
spatial resolution,
time resolution,
high particle flux capability,
radiation tolerance,
gamma sensitivity, background discrimination,
time stability, and ageing.

In the next paragraphs we summarise the contributions concerning the thermal neutron detection according to what in our believe are the most important discussion topics.

\subsection{Simulations}
Deterministic and Monte Carlo (MC) simulations are valuable and \emph{costless} tools to address the research toward new neutron detectors.
%Deterministic methods provide more exact solutions of approximate models, whereas Monte Carlo methods provide approximate solutions of more exact models (stochastic events)
MC methods are suited to study stochastic processes, and in particular the transport of radiation through matter.
The accuracy and reliability of MC predictions depends on the model and the input data libraries.
Therefore, simple experiments must be setup to benchmark their performances.

In the contribution of Lina Quintieri~\cite{Quintieri_talk}, the thickness of LiF and $^{10}$B coating on a single layer and a multi layer cathode was optimised using GEANT4~\cite{GEANT4} and FLUKA~\cite{FLUKA} simulation packages for the GEM \emph{side-on}~\cite{SideOn} detector.
The two simulations give similar results, and the $^{10}$B behaviour is well in agreement with the experimental data.
The optimal thickness of the $^{10}$B is found to be 1.8~$\mu$m for five layers of $^{10}$B converter.

%The conversion efficiency and the secondary particle yield is studied as a function of the thickness of a single layer of LiF.
%Comparison with the data is not yet possible, but GEANT4 and FLUKA simulations give comparable results.

\begin{figure}
\centering
\includegraphics[width=0.8\linewidth]{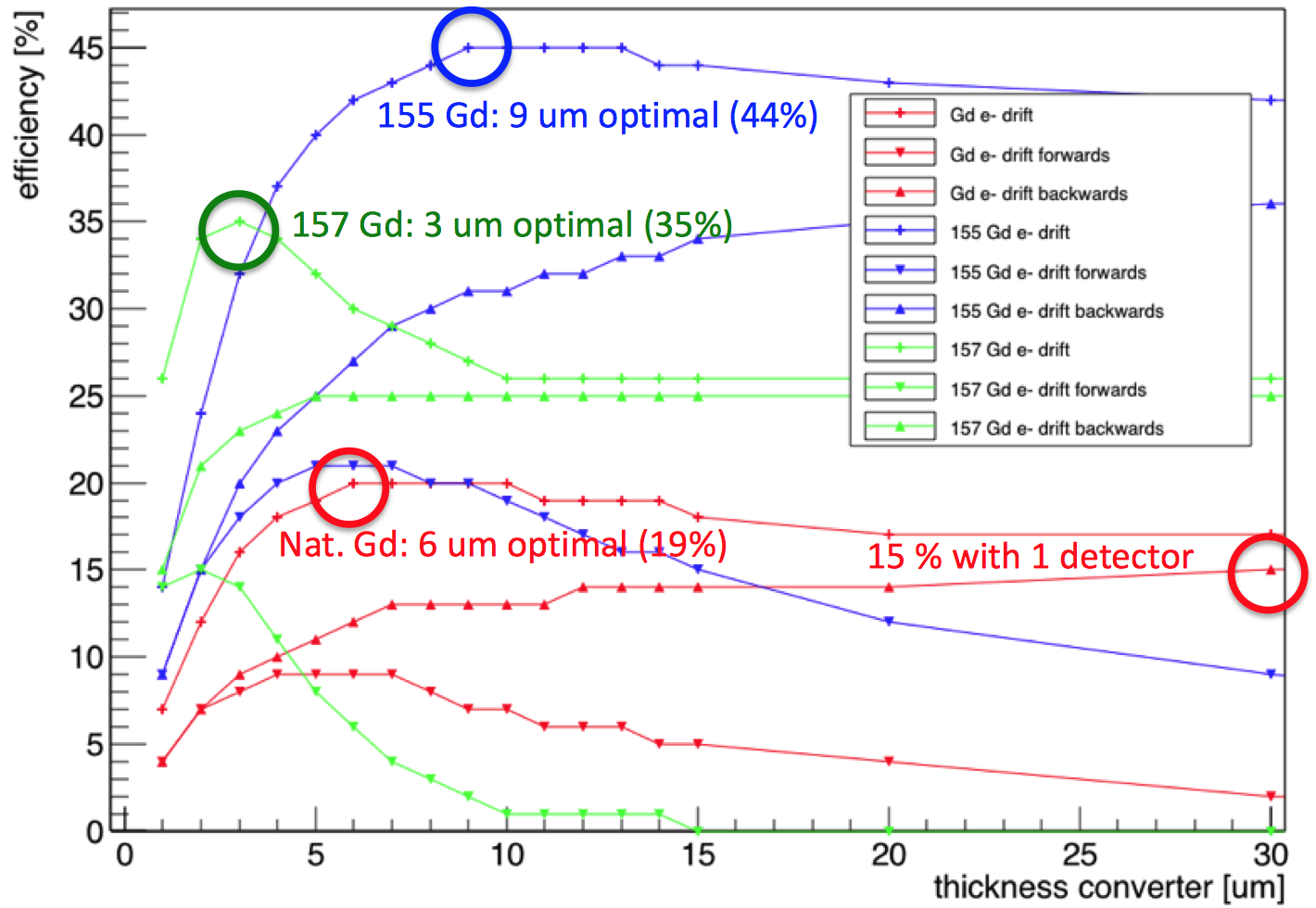}
\caption{Simulated detection efficiency for different Gd-based converters.}
\label{fig:GdSim}
\end{figure}
$^{6}$Li and $^{10}$B are not the only converter material studied in simulations.
Dorothea Pfeiffer~\cite{Pfeiffer_talk} simulated with GEANT4 the performance of cathodes singly and doubly coated with natural Gadolinium, $^{155}$Gd, $^{157}$Gd, Gd$_2$O$_3$ and enriched Gd$_2$O$_3$.
As shown in figure~\ref{fig:GdSim}, 9~$\mu$m of $^{155}$Gd has a conversion efficiency up to a very promising 44\% for 25~meV neutrons.
The simulation considers also the energy distribution of the conversion electrons in several thicknesses of gas.
GEANT4 is interfaced with GARFIELD++~\cite{GARFIELD++} for the simulation of the transport and the amplification of the primary ionisation electrons.

\subsection{Coating}
Solid converters deposited on adequate substrates are the most suitable solutions for MPGD based neutron detectors.
B$_4$C is one of the most popular materials, that can be enriched with $^{10}$B.
In order to further increase the detection efficiency, efforts are ongoing to deposit pure metal boron films~\cite{Pietropaolo_talk}.
A collaboration between ENEA, INFN, Columbus Superconductors~SpA~\cite{Columbus} and RHP Technology~\cite{RHP} resulted in tests performed with an evaporation system at ENEA Frascati, a radio-frequency sputtering at ENEA Casaccia, and High Power Impulse Magnetron Sputtering at INFN LNF.
The coatings were successfully used in the multi-layer cathode of the GEM \emph{side-on} detector, to reach efficiencies as high as~31\% for 5~meV neutrons~\cite{Claps}.

Some of the ESS neutron scattering instruments will require very large converter surfaces.
The more common $^{10}$B$_4$C, in this case, seems preferable.
Nevertheless, the ability to mass-produce with the acceptable quality and speed hundreds of m$^2$ must be proven~\cite{Robinson_talk}.
Since the Summer 2014, the ESS detector coatings workshop in Link\"{o}ping is equipped with a 4 magnetrons sputtering machine from CemeCon~\cite{Cemecon}, able to coat at the same time several large ($50\times10$cm$^2$) targets.
27000 blades (more than 100~m$^2$) for a full IN5 sector demonstrator~\cite{Khaplanov} were coated with $^{10}$B$_4$C in 45~days, proving the mass production capabilities of the ESS coatings workshop in Link\"{o}ping.
$10^{14}$~n/cm$^2$ on a 1~$\mu$m thick film are proven not to influence on adhesion, composition, morphology, and structure.
B$_4$C can be properly transferred to several substrates like aluminium, stainless steel, alumina, silicon, copper and Kapton.
Difficult substrate, where the adhesion is not optimal and more R\&D is needed are glass, Teflon, nickel and MgO.
%In addition, R\&D efforts are ongoing to transfer a stable gadolinium compound on metals.

\subsection{Efficiency}
\begin{figure}
\centering
\includegraphics[width=0.8\linewidth]{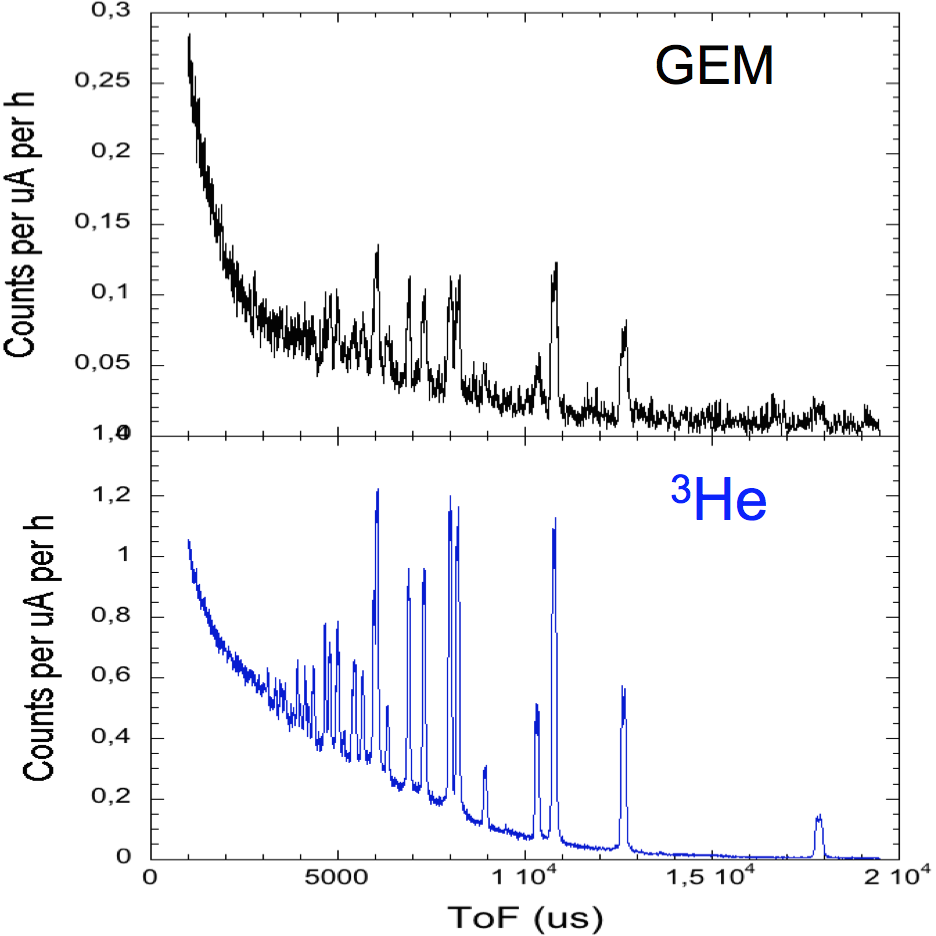}
\caption{Comparison of a diffractogram of a bronze sample from a boron-GEM and an $^3$He tube. The total measurement time is about 20~h. The active area of the GEM detector is about $5\times10$~cm$^2$, while the active area of the $^3$He tube is about $2.5\times15$~cm$^2$.}
\label{fig:tof}
\end{figure}
The first diffractogram recorded with a GEM based detector was reported~\cite{Croci_talk}.
A \emph{low efficiency} B$_4$C coated aluminium cathode was used for this measurement at INES-ISIS.
The efficiency was increased of about a factor of three with a $^{10}$B enriched boron carbide cathode.
The diffractogram of a bronze sample is shown in figure~\ref{fig:tof} in comparison to the one recorded with an $^3$He tube.
The poorer peak resolution is mainly attributed to the lack of a neutron collimation in front of the detector.
The resolution was improved summing the spectra from pads that lie on the same Debey-Scherrer cone.
In order to further increase the detection efficiency, new cathode geometries are under study.
In particular, the new BAND-GEM detector, with B$_4$C-coated \emph{lamellas} that resample the field shaping electrodes of a TPCs, was tested at the RD2D beam-line at IFE, demonstrating about 15\% efficiency for neutrons with energies of 34.5~meV.

%Eleni Aza
%increase efficiency
%different geometry gem side on
%5 x 5 cm2 active area
%3 x 6 mm2 pads (128)
%Beam profile at EAR1 of nTOF with side-on detector

\begin{figure}
\centering
\includegraphics[width=0.8\linewidth]{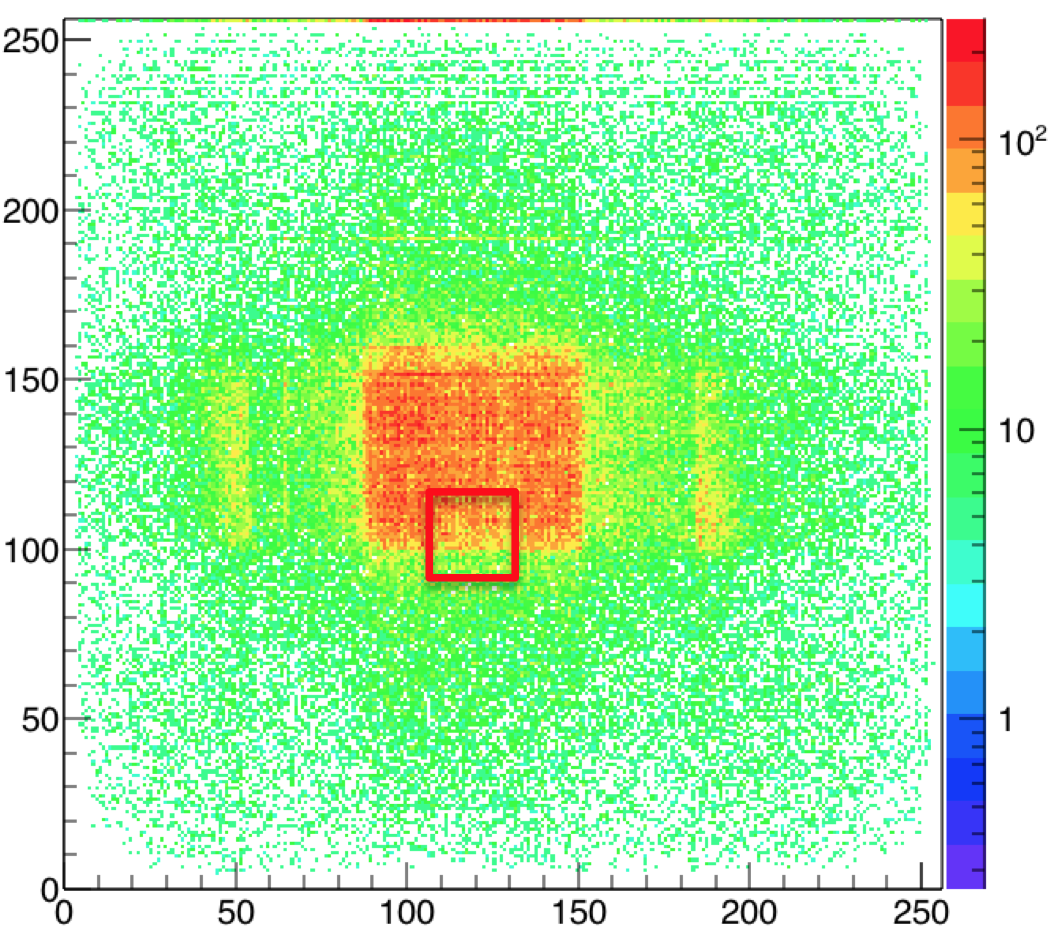}
\caption{Beam profile from a Gd-GEM at RD2D beam-line at IFE. The band structure on the sides are intentional dis-homogeneities in the cathode.}
\label{fig:scatter}
\end{figure}
The Neutron Macromolecular Crystallography (NMX) instrument at ESS requires detectors with efficiency larger than~20\%, spatial resolution of about 200~$\mu$m, capabilities to cope with very high local neutron fluxes and a wide range of incident neutron angles.
This seems to imply that the neutron capture must occur in a single conversion layer.
Li and B based converters are not enough efficient, instead gadolinium, despite other complications, seems promising.
For the first time, a triple GEM was coupled to a 250~$\mu$m natural Gd cathode and exposed to a neutron beam~\cite{Pfeiffer_talk} at RD2D beam-line at IFE.
The detection efficiency was estimated to be about~10\%, smaller than the simulated one, but with large systematic uncertainties.
The profile of a square beam is shown in figure~\ref{fig:scatter}.
The spatial resolution is estimated from the sharp edge marked in red.
The TPC analysis described in the next paragraph improves the spatial resolution by a factor of three.
The spatial resolution was estimated to be of the order of 1~mm, dominated by the neutron scattering on the materials in front of the detector.
In spite of the fact that gadolinium-based detectors are more sensitive to gamma, the signal to background ratio in the RD2D beam-line conditions was 100:1.
Boron-based detectors, depending on the gain, can reach gamma rejections of the order of $10^-7$~\cite{Khaplanov:2013}.

%M. Sabaté-Gilarte (low efficiency)
%$n_TOF$ spallation source at cern 
%Microbulk mm beam monitoring for ntof
%transparent low efficient
%Minimal beam perturbation
%Cover wide range of energy
%High geometrical efficiency
%Minimize induced background by the device itself
%Flux measurement beam profile
%10B 6Li 235U as converter
%used also for Fission cross-section measurements

\subsection{Space resolution}
\begin{figure}
\centering
\includegraphics[width=0.8\linewidth]{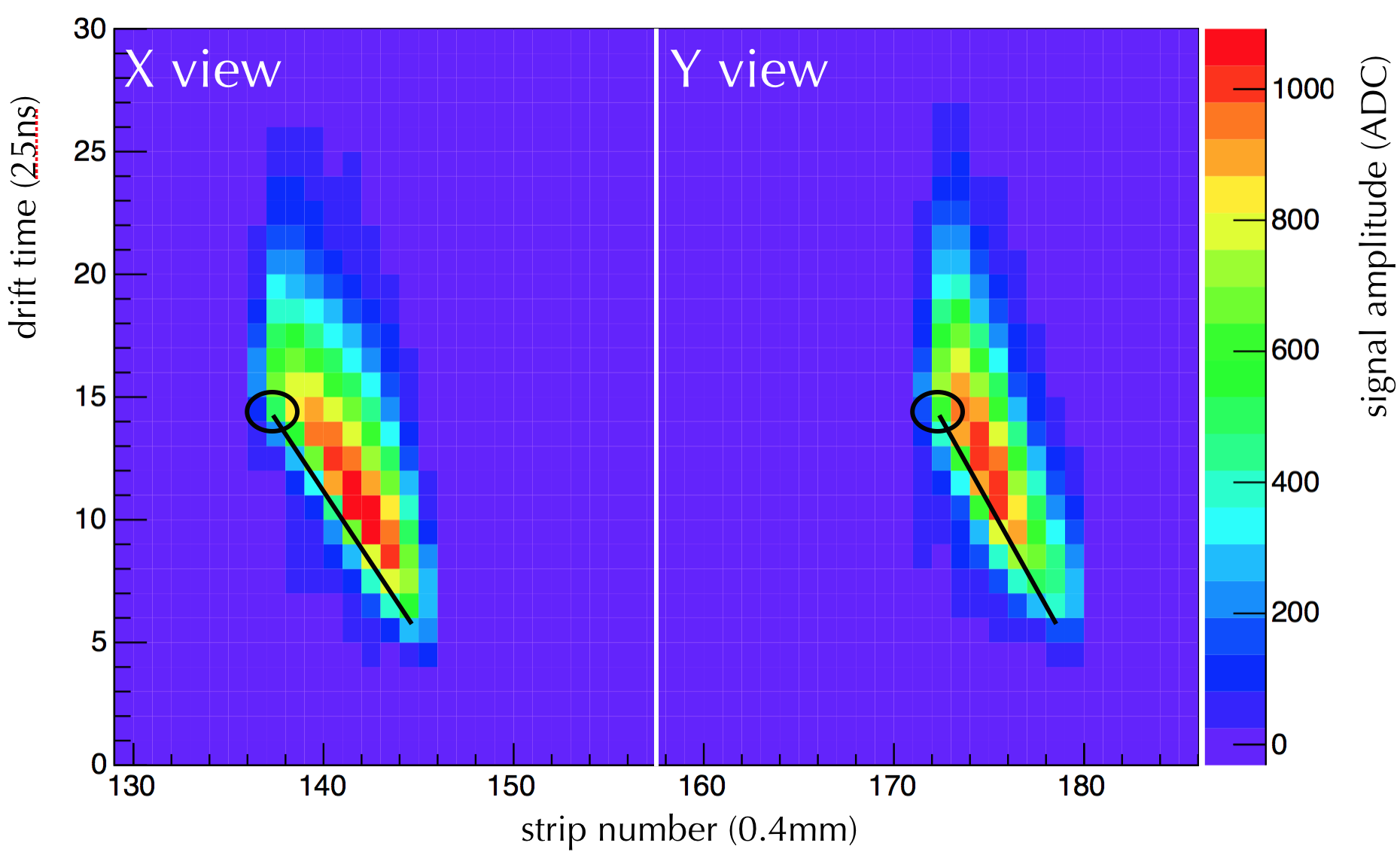}
\caption{Two views of a $^7$Li or $\alpha$ track from a neutron conversion in $^{10}$B. The point marked with the circle is the extrapolated point of the neutron capture.}
\label{fig:events}
\end{figure}
In the case of $^{10}$B and Gd, the typical range of the secondary charged particles in gas is several millimetres, apparently in contrast to some ESS instruments requirements of sub-millimetre position resolution.
A single GEM detector with a $^{10}$B$_4$C coated cathode and an anode segmented into two orthogonal sets of 400~$\mu$m wide strips demonstrated position resolutions on the two views better than 200~$\mu$m~\cite{Resnati_talk, Pfeiffer}.
The detector is used as an imaging device, profiting from an analysis based on the concept of the Time Projection Chamber (TPC).
The neutron capture point in the converter is extrapolated on an event by event basis from the $\alpha$ and $^7$Li tracks (see figure~\ref{fig:events}).
With a resistive MicroMeGas, similar results are obtained on one strip view.
The other one is affected by the electrons evacuation through the resistive layer, and the signal shapes are not suitable for the TPC analysis.
With respect to the \emph{center of gravity} approach, the TPC analysis improves the position resolution of a factor of~5.
Since the analysis is independent of the detector details, it can be exploited in several occasions.
A similar analysis significantly improves the position resolution on Gd-based detectors~\cite{Pfeiffer_talk}.
Given the curly trajectories of tens keV electrons from the Gd neutron conversion, a refined analysis can improve even more significantly the Gd-GEM performance.

\begin{figure}
\centering
\includegraphics[width=0.8\linewidth]{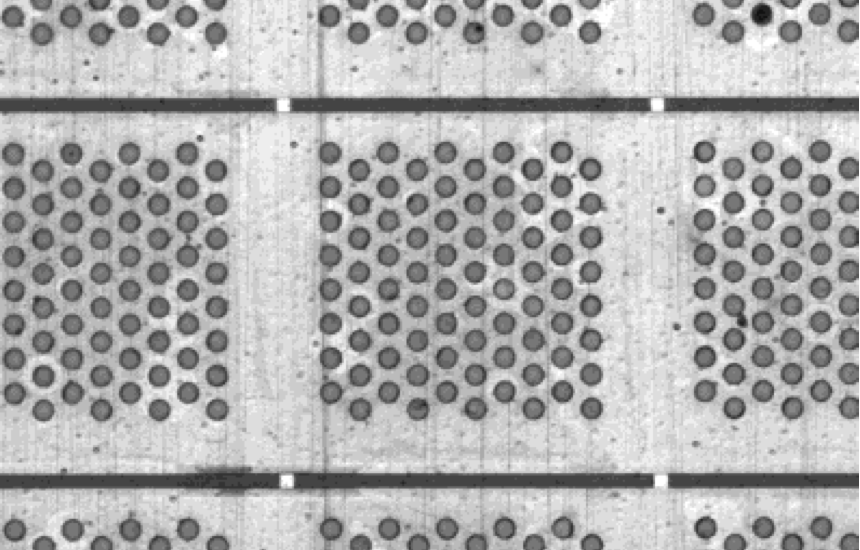}
\caption{Closeup of the holes of the micro-bulk MicroMeGas. The electrode is segmented into strips. In transparency one can notice the orthogonal segmentation of the other electrode.}
\label{fig:strips}
\end{figure}
There are circumstances where severe constraints drive the entire design of the detector.
Extremely low \emph{material budgets} are necessary when the detector is installed in the neutron beam as profiler~\cite{Diakaki_talk}.
Detection efficiency is clearly not an issue, position resolutions of about 1~mm, easily achievable in other conditions, become a challenge when almost no metal is allowed in the detector.
A micro-bulk MicroMeGas with orthogonally segmented anode and \emph{mesh} is tested in a neutron beam at GELINA FACILITY IRMM-GEEL~\cite{GEEL}.
The 1~mm wide strips on the anode and the mesh provide two dimensional readout, without an addition of a dedicated 2D~anode.
The picture in figure~\ref{fig:strips} shows a closeup of the segmented electrode of the MicroMeGas.
In transparency the orthogonal strips can be noticed.

\subsection{Rate capability}
MPGDs were invented to improve the position resolution and the particle flux capability of the MWPCs.
In particular, GEMs at gains of few thousands are known to well cope with fluxes of the order of 1~MHz/mm$^2$ of minimum ionising particles crossing 3~mm drift region.
The high neutron flux capability of a \emph{standard} triple GEM with the cathode coated with B$_4$C was tested at the G3-2 irradiation station at the ORPHEE reactor (LLB-Saclay).
The saturation of the measured rate above 10~MHz/cm$^2$ is due to the electronics dead time, while the detector gain is independent of the particle flux in these ranges~\cite{Croci_talk}.

\section{Fast Neutron Detectors}
In order to detect fast neutrons (E $>$ 1~keV), different techniques have been used, but most of them are based on the usage of hydrogenated materials that can either moderate fast neutrons down to the thermal levels, or produce charged particles by elastic scattering that can ionise the active gas of different kind of MPGDs.
Both GEM and $\mu$Megas detectors have been developed to detect fast neutrons in several applications linked to beam monitoring, neutron cross-section measuring and total dose estimations. 
Applications related both to spallation sources, fusion reactors and medical field profited from the introduction of Micro-patterned gaseous detector to improve instrumentation performance.

The properties of MPGD based fast neutron beam monitors can be summarised as follows:
\begin{itemize}
\item Neutrons are converted through nuclear reactions on
\begin{itemize}
\item A hydrogenated material (such as polyethylene or polypropylene),
\item A material containing boron or lithium (after thermalisation),
\item A uranium ($^{235}$U) sheet,
\item A gas containing helium or hydrogen.
\end{itemize}
\item Minimal perturbation of neutron beam,
\item Minimal introduction of background,
\item Capability of reconstructing the 2D shape of the beam,
\item Capability of on-line measuring the neutron flux also up to very high rates,
\item Efficiency from 10$^{-2}$ to 10$^{-7}$,
\item Space resolution from few mm to cm,
\item Time resolution of  few ns,
\item Gamma ray rejection from 10$^{-6}$ to 10$^{-7}$.
\end{itemize}

Micromegas detectors coupled with $^{10}$B or $^{235}$U cathodes have been used at nTOF~\cite{nTOF} since 2001 to monitor the beam profile.
The device, with an active area of $6\times6$~cm$^2$ is composed by several $\mu$bulk Micromegas inserted in an aluminium gas chamber together with converter samples. 

\begin{figure}[h]
\minipage{0.5\textwidth}
\centering
\includegraphics[scale=0.4]{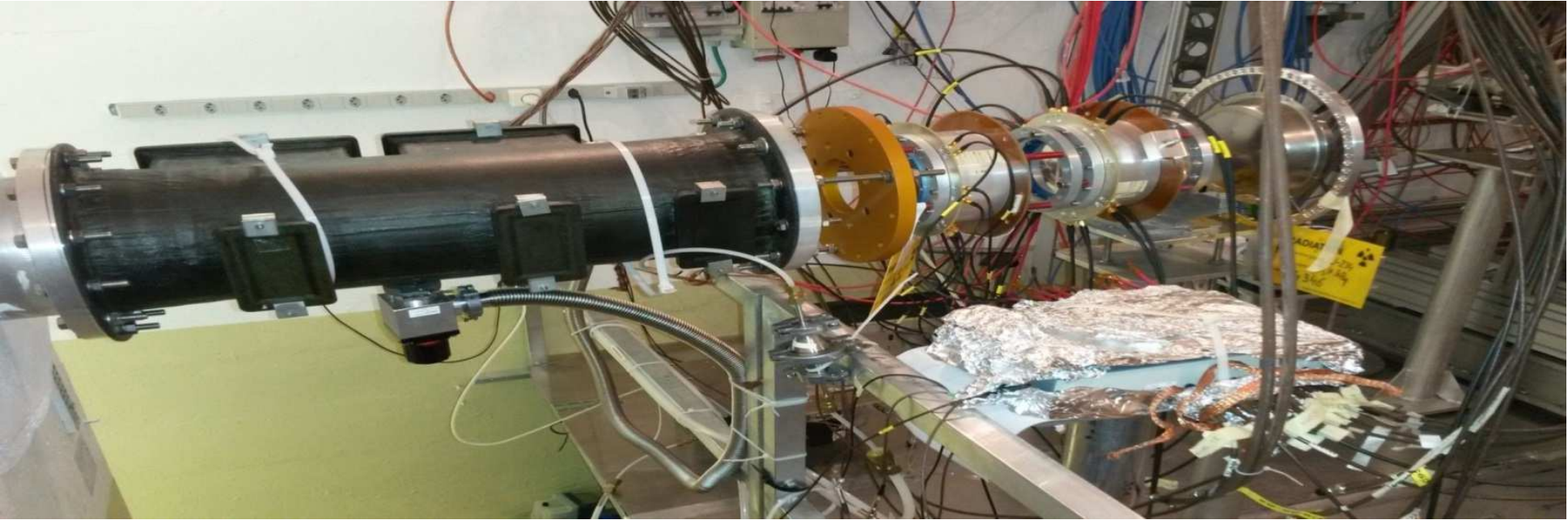}
\caption{$\mu$Megas installed on the nTOF beam pipe.}
\label{mm1}
\endminipage
\hfill
\minipage{0.5\textwidth}
\centering
\includegraphics[scale=0.5]{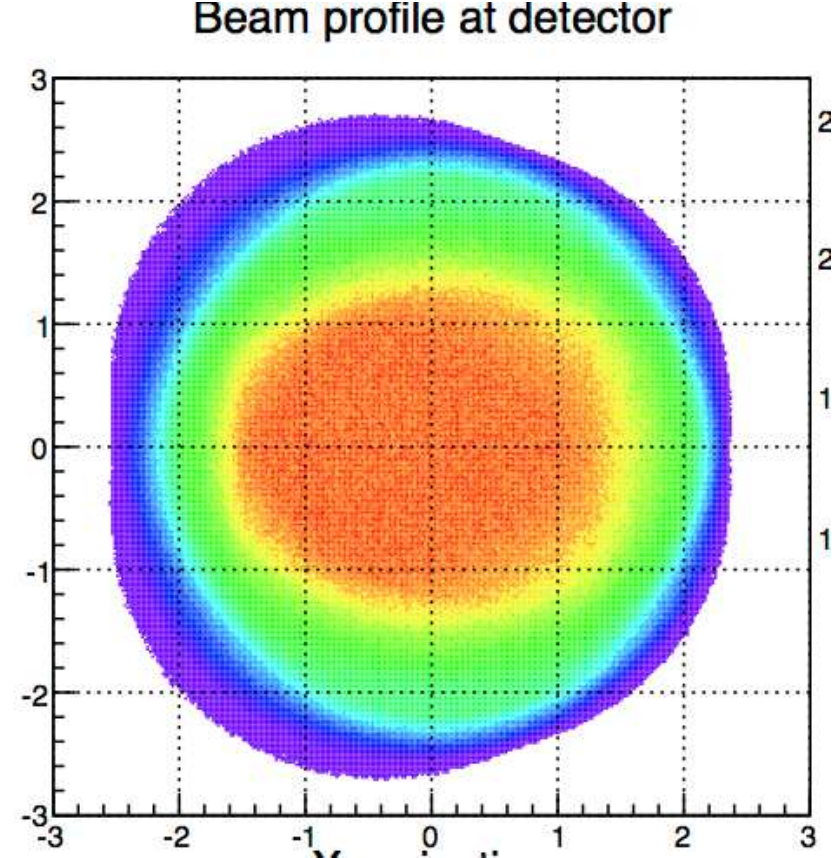}
\caption{Beam profile reconstructed through X-Y $\mu$Megas.}
\label{mm2}
\endminipage
\end{figure}

New tracking X-Y Micromegas beam monitors based on the $\mu$bulk technique have also been developed giving the possibility to improve space resolution through the track reconstruction algorithm.
These detectors (see Figure \ref{mm1}) installed on the nTOF beam pipe are used to measure both the neutron beam profile (see Figure \ref{mm2}) and the neutron flux.
These detectors have a very low material budget and thus do not introduce any perturbation in the beam itself~\cite{Diakaki_talk}.

\begin{figure}
\centering
\includegraphics[scale=0.5]{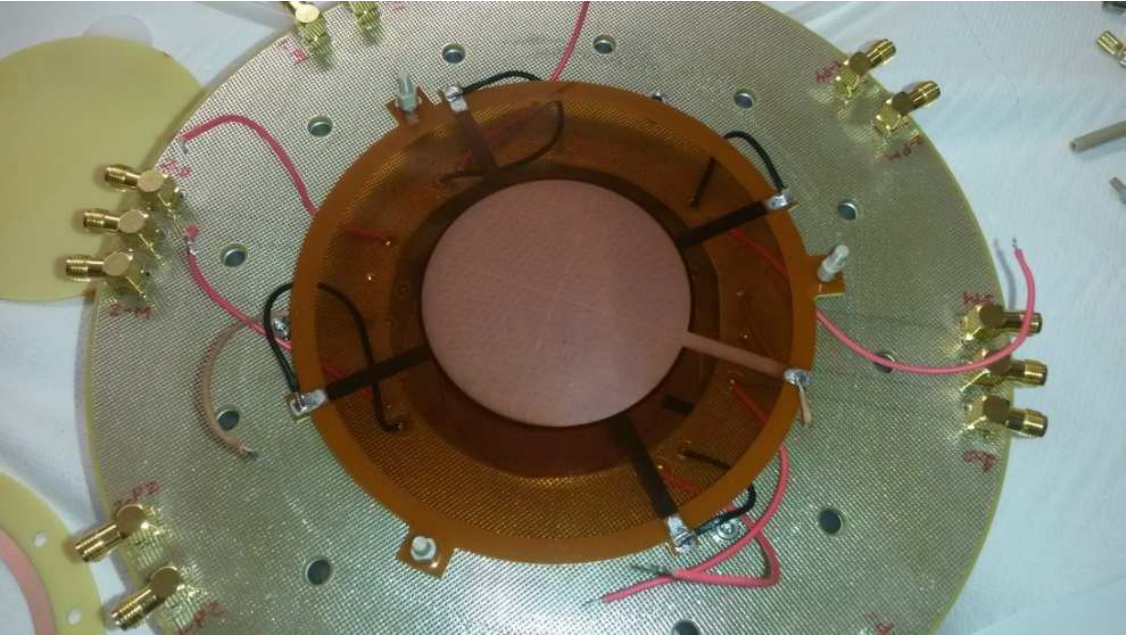}
\caption{Picture of the 4 pads Micromegas.}
\label{mm3}
\end{figure}

\begin{figure}
\centering
\includegraphics[scale=0.2]{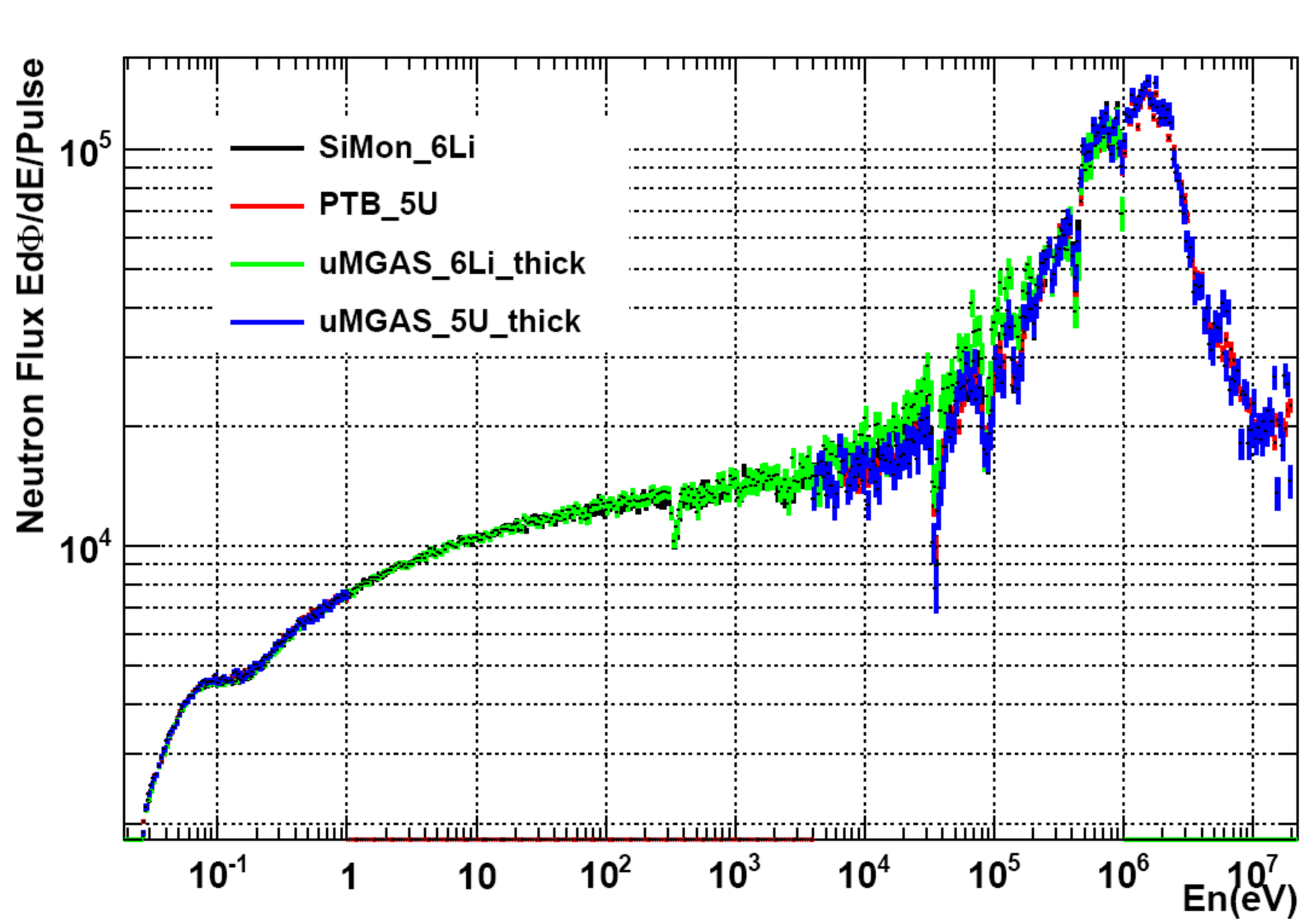}
\caption{nTOF neutron flux reconstructed with the 4 pads Micromegas}
\label{mm4}
\end{figure}

This kind of detector have been used also to measure fission cross-sections (n,f) and (n,charged particles) of materials like $^{235}$U, $^{240}$Pu, $^{242}$Pu and $^{33}$S. 
The detector has been recently improved with the use of a 4 pads-anode instead of a standard X-Y strips that allows reducing the capacitance and increasing the counting rate capability and the S/N ratio (see Figure \ref{mm3}).
This kind of chamber has been already used to measure the neutron flux (see Figure \ref{mm4}).
The active area of the device is still a circle with diameter of 6~cm~\cite{talk_mm2}.

Other Micromegas devices have been developed as portable and directional fast neutron detectors.
This devices are able to detect neutrons from few keV up to several MeV by exploiting He based gas mixture and measuring the recoil atoms produced by a neutron interaction.
The active area covered by this portable devices is 100~cm$^2$~\cite{talk_mm3}

\begin{figure}
\centering\includegraphics[width=0.95\linewidth]{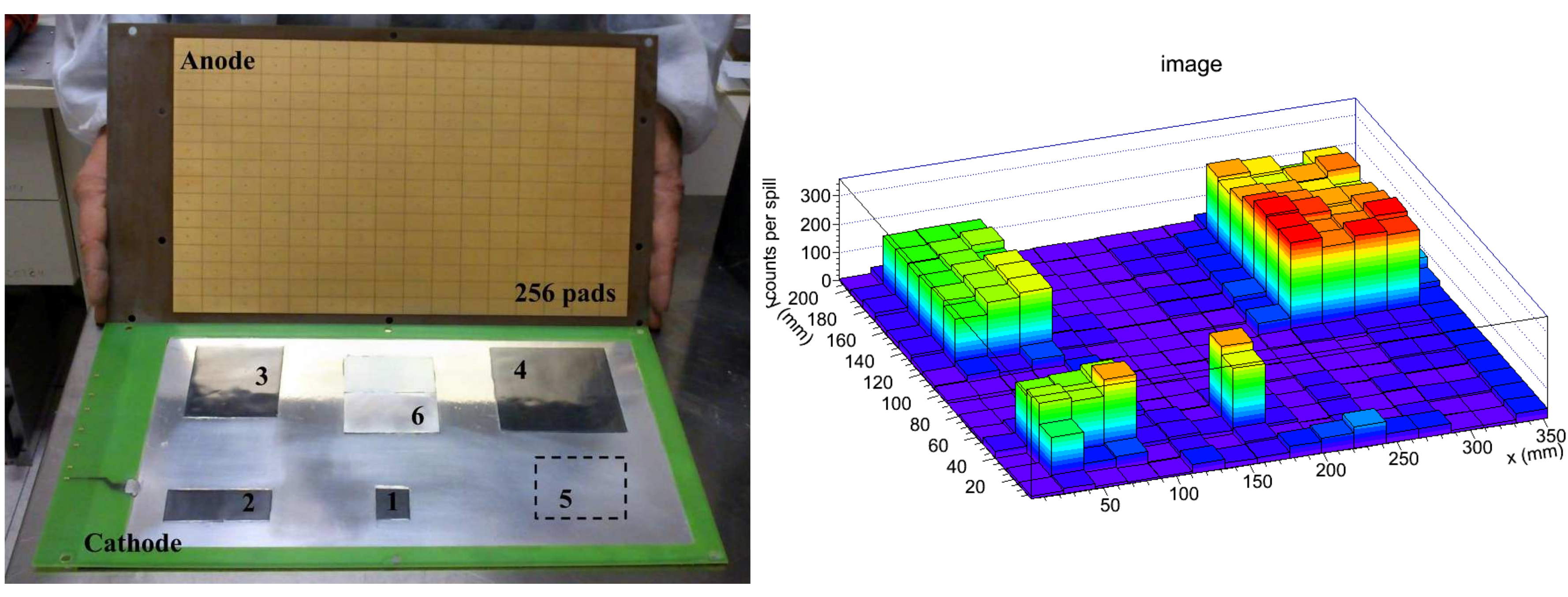}
\caption{GEM-based neutron spectrometer and its event display.}
\label{gem1}
\end{figure}

GEM-based fast neutron detectors have also been presented for applications related to spallation sources, to diagnostic of neutral beam injector facilities for future fusion reactors and to medical applications.
Several GEM detectors of different sizes (from $10\times10$~cm$^2$ up to $20\times35$~cm$^2$~\cite{gab1, gab2} ) have been used to measure the nTOF beam properties (both energy spectrum and profile).
These chambers are provided with different kind of cathodes (both polyethylene, polypropylene and borated) and have different sizes.
2D beam images as well as neutron energy spectra have been measured by exploiting the neutron time of flight technique and show that the thermal and fast neutron components of the beam have different shapes.
The space resolution achievable with these devices depends upon the pad dimension and ranges from 1~mm up to 10~mm.
Recent development is the realisation of a medium size area GEM based spectrometer equipped with a cathode subdivided in different areas that are sensitive to different neutron energies (see Figure~\ref{gem1}.
Areas have all different shapes and some of them are borated and if covered with polyethylene slabs work like a planar Bonner-sphere~\cite{talk_gem1}.

\begin{figure}
\centering
\includegraphics[scale=0.5]{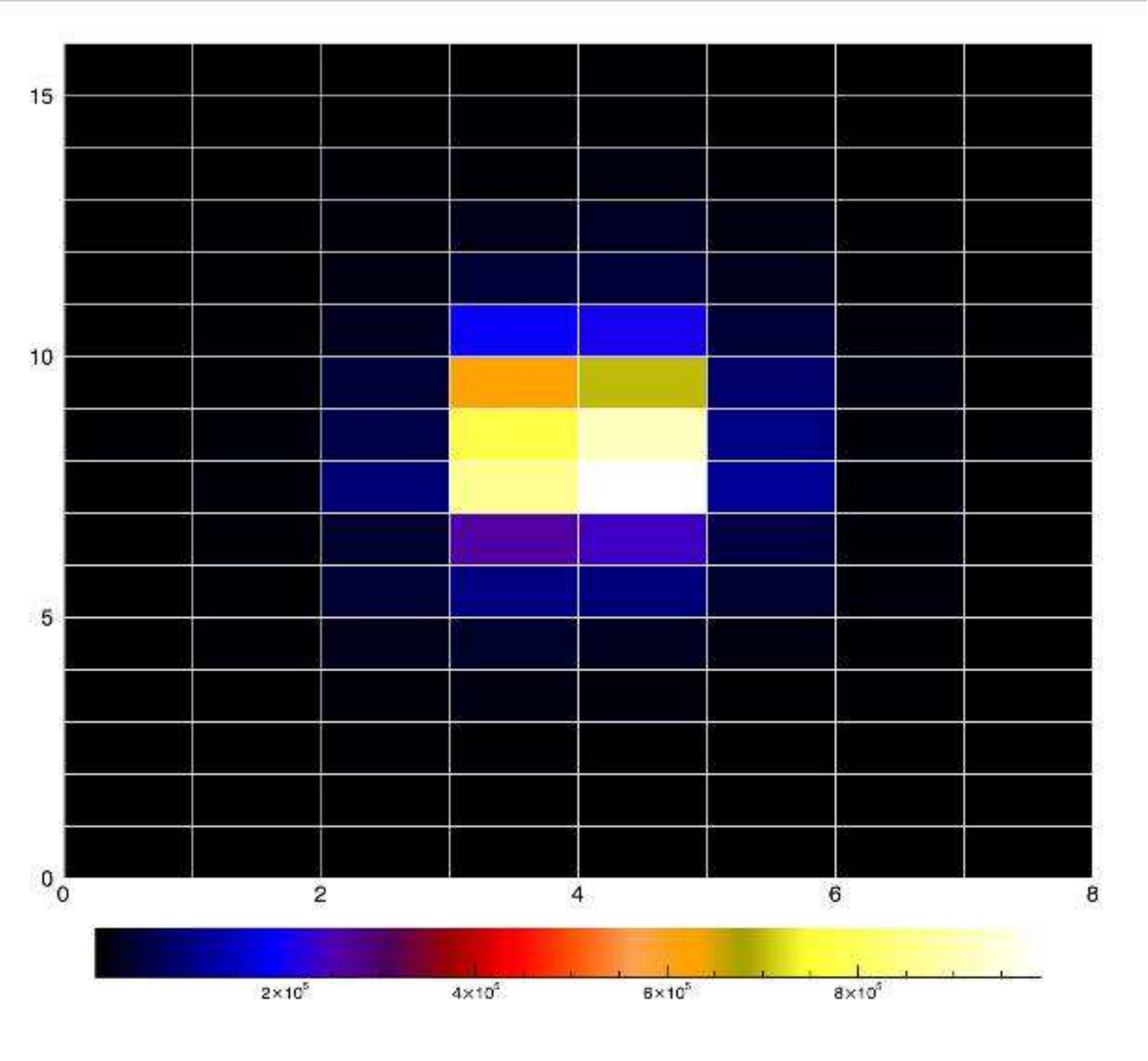}
\caption{Beam Profile reconstruction using the $35\times20$~cm$^2$ nGEM detector for SPIDER. The spatial resolution achieved in this case is about 10~mm due to the pad size (13~mm~$\times$~22~mm).}
\label{gem2}
\end{figure}

Another medium-size ($35\times20$~cm$^2$) GEM detector has been developed for applications related to thermonuclear fusion %~\cite{nGEM-spider}
in a facility called SPIDER %~\cite{nGEM-spider}
that represents the prototype of Neutral beam injector for ITER.
The neutron production due to fusion reactions between beam deuterons and deuterons implanted in the SPIDER dump will be a few times 10$^{15}$ neutrons per SPIDER pulse.
The neutron source intensity is suitable for diagnostic applications.
A neutron diagnostic was designed in order to provide the map of the beam intensity with a spatial resolution approaching the size of individual beamlets is described.
The proposed detection system is called Close-contact Neutron Emission Surface Mapping (CNESM) and it is placed right behind the beam dump as close as possible to the neutron emitting surface (around 30~cm).
The detectors employed in this diagnostic system are nGEM chambers~\cite{nGEM}.
This device was tested at the ISIS spallation source and shows same performance as small area ($10\times10$~cm$^2$) detectors~\cite{gab2}.
Figure~\ref{gem2} shows the reconstructed ISIS beam profile.
Its uniformity of response all over the active area is higher than~90\% proving that larger area GEM based fast neutron detectors can be realised.
This kind of chambers can be also used as beam monitors for fast neutron lines at spallation source (like ChipIR at ISIS) or for other irradiation lines~\cite{talk_gem2}.. 

\begin{figure}
\centering
\includegraphics[scale=0.5]{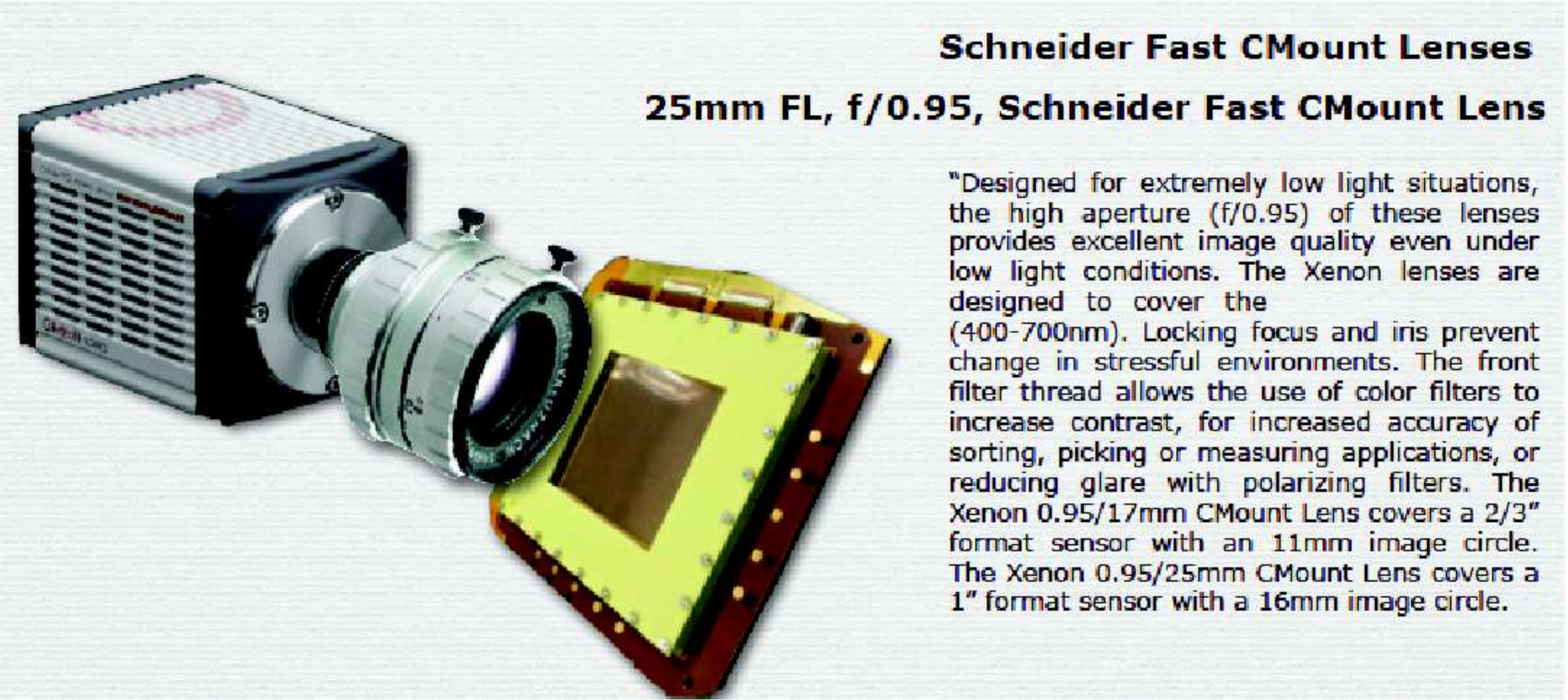}
\caption{Schematic of the MONDO device.}
\label{gem4}
\end{figure}

Finally a medical application related to hadron-therapy was presented. 
The aim is the measurement of the neutron flux due to the proton interaction within the patient body during the treatment: due to spallation reaction in the body, a very high neutron flux: For example for a proton beam of 172 MeV/u, an average differential neutron flux $\Phi_E$ of 10$^4$ $[(MeV/u \times cm^2)^{-1} Gy^{-1}]$ is produced.
A GEM based device (MONDO) coupled with plastic scintillators and with a CMOS camera is being designed to measure the neutron flux and the direction of the produced neutrons.
This device convert neutrons from 20 to 200~MeV into charged particles that produce scintillation lights in the fibre that is subsequently converted into photo-electrons by CsI coated GEMs.
The photo-electrons are then further multiplied and generate secondary light that can be detected by a very high resolution commercial CMOS photon camera.
Figure~\ref{gem4} shows a schematic of the system and gives more details about the camera.
One of the most important features is that with such a scheme the electronics is kept far away from the neutron beam thus increasing its survival time.
Fluka simulations show that this system gives the possibility to measure the neutron flux and to understand the direction of the neutrons \cite{talk_gem3}.

\begin{figure}
\centering
\includegraphics[scale=0.5]{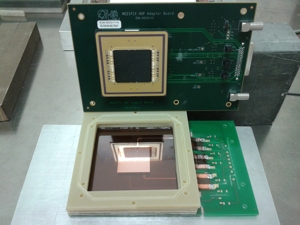}
\caption{Image of a GEMPix detector.}
\label{fig:GEMPix}
\end{figure}

Coupling MPGD detectors to CMOS read-out gives the possibility to realise very high resolution detectors.
A quad-Timepix~2 chip was used as anodic readout of a Triple GEM detector and this device was called GEMPIX~\cite{Murtas}.
The Timepix~2 chip has an active area of $1.4\times1.4$~cm$^2$ and is composed by $254\times254$ pixels each with an area of 55~$\mu$m~$\times$~55~$\mu$m.
The GEMPix detector (picture in figure~\ref{fig:GEMPix}) has an active area of $28\times28$~mm$^2$ since it is read-out by a quad-Timepix (4 chips).
Such a high spatial resolution implies the capability of recognising every single interacted track giving the possibility to operate the particle identification by exploiting the track morphology.
This opens the way to a so called microscopic-analysis that improves the reconstruction of macroscopic features like beam profile.
Capability of standard GEMPix detector under MIPs irradiation were presented as well as a preliminary measurement in a mixed neutron field.
By exploiting either the TOT or the TOA both 2D maps as well as energy spectra were reconstructed.
This kind of chip can be also used coupled to standard silicon detectors that proved to be able to reconstruct the nTOF beam profile as well as to measure the neutron flux and the neutron energy spectrum \cite{talk_gem4}.
All the results shown during the workshop prove that micro pattern Gaseous Detectors offer very high performances for fast neutron detection in different applications field giving the possibility to 
improve already existing technology or to explore new areas.

\section{Electronics}
Several front end electronics have been used coupled to MPDG-based neutron detectors but, at the moment, they are all derived or adapted from electronics developed either for silicon or Multi Wire Proportional Chambers, so none of them has been especially developed for MPGDs.
Nevertheless, some of them have proven to give reliable results also when applied to neutron detectors based on MPGDs.
Table~\ref{table} shows some of the electronics chips that have been recently used coupled with MPGDs. 
Standard requirement for electronics to be used for neutron measurement is radiation hardness, since it is known that neutrons can be harmful for chips since they induce Single Event Upsets (SEU) that can lead even to the disruption of the chip itself.
The RD51 collaboration~\cite{MPGD} has recently developed a system called SRS (Scalable Readout System) that is meant to be a base DAQ to be interfaced with different kind of Front-End Chips.
Up to now the only chip that has been used with the SRS is the APV25 analog chip, that was also used for the read-out of $\mu$TPC thermal neutron chamber and of Gadolinium GEM detectors (both presented during this workshop \cite{Pfeiffer_talk, Resnati_talk}.
Although APV25 coupled to SRS does not give the possibility to acquire at high rates (it is limited at order of kHz), its feature of being bunched in 25 ns gates can be beneficial for very precise ToF measurements of fast neutrons.
For example, if it is triggered by the proton beam T0 (that is when the beam impinges on a spallation target) it can be used to improve the fast neutron energy spectrum measurements that have been shown during the workshop.
Other chips that will be coupled to the SRS in the near future are the GEMROK and the VMM2 (see the table for details).

\begin{sidewaystable}[h]
\small
\centering
\caption{List of some of the possible front-end chips to be used with neutron MPGDS}
\label{table}
\begin{tabular}{|l|l|l|l|l|l|l|l|l|l|l|l|}
\hline
Name                                                                  & Costumer                                                                        & Maker                                                                     & Type                                                                                                     & Chan & Shaping                                                        & \begin{tabular}[c]{@{}l@{}}Det\\ Cap\end{tabular} & \begin{tabular}[c]{@{}l@{}}Sens\\ (mV/fC)\end{tabular} & Noise                                                         & \begin{tabular}[c]{@{}l@{}}Time\\ stamp\end{tabular}         & \begin{tabular}[c]{@{}l@{}}Max\\ Rate\end{tabular}          & \begin{tabular}[c]{@{}l@{}}Rad\\ Hard\end{tabular}       \\ \hline
Carioca                                                               & \begin{tabular}[c]{@{}l@{}}LhCb \\ Muon\\ (MWPC)\end{tabular}                   & \begin{tabular}[c]{@{}l@{}}CERN\\ INFN\end{tabular}                       & \begin{tabular}[c]{@{}l@{}}Discr.\\ A-S-D\\ MWPC\end{tabular}                                            & 8    & 15 ns                                                          & \textless220pF                                    & 10                                                     & \begin{tabular}[c]{@{}l@{}}600e\\ (100 pF)\end{tabular}       & N/A                                                          & \begin{tabular}[c]{@{}l@{}}10\\ MHz/ch\end{tabular}         & \begin{tabular}[c]{@{}l@{}}Up\\ to 1\\ Mrad\end{tabular} \\ \hline
\begin{tabular}[c]{@{}l@{}}GEMROK \\ (MSGCROK\\ derivative)\end{tabular} & \begin{tabular}[c]{@{}l@{}}DETNI \\ Neutrons\end{tabular}                       & \begin{tabular}[c]{@{}l@{}}Krakow\\ Heidelberg \\ HMI Berlin\end{tabular} & \begin{tabular}[c]{@{}l@{}}A-2S-PD-A-T\\ MSGC \\ GEM\\ Peak\\ Analog\\ Pipe\\ Prompt\\ disc TS\end{tabular} & 32   & \begin{tabular}[c]{@{}l@{}}Dual \\ 25/85 ns\end{tabular}       & aim 25 pF                                         & 2                                                      & \begin{tabular}[c]{@{}l@{}}aim\\ 200 e\end{tabular}           & \begin{tabular}[c]{@{}l@{}}12-14\\ bit  \\ 8 ns\end{tabular} & N/A                                                         & N/A                                                      \\ \hline
APV25                                                                 & \begin{tabular}[c]{@{}l@{}}CMS \\ Tracker \\ (Si Strip)\end{tabular}            & RAL                                                                       & \begin{tabular}[c]{@{}l@{}}Analog\\ Pipe\\ A-S-BUF\end{tabular}                                          & 128  & \begin{tabular}[c]{@{}l@{}}50 ns +\\ fast\\ decon\end{tabular} & low                                               & 1 mA/fC                                                & \begin{tabular}[c]{@{}l@{}}2000 e\\ 430+61\\ /pF\end{tabular} & LHC                                                          & \begin{tabular}[c]{@{}l@{}}100 Hz\\ with\\ SRS\end{tabular} &                                                          \\ \hline
VFAT                                                                  & CMS                                                                             & \begin{tabular}[c]{@{}l@{}}CEA\\ INFN\\ CERN\end{tabular}                 & \begin{tabular}[c]{@{}l@{}}Disc \\ Pipe \\ A-S-\\ D-BUF\end{tabular}                                     & 128  & 25-400 ns                                                      & 5-80 pF                                           & 1-50                                                   & \begin{tabular}[c]{@{}l@{}}1000 +\\ 40/pF\end{tabular}        & LHC                                                          & N/A                                                         & N/A                                                      \\ \hline
CBC                                                                   & CMS Si Strip                                                                    & RAL                                                                       & \begin{tabular}[c]{@{}l@{}}A-S-D\\ Analog\\ prot\end{tabular}                                            & 128  & 20 ns                                                          & 3-6 pF                                            & 50                                                     & \begin{tabular}[c]{@{}l@{}} 1000 e \\ at 10pF\end{tabular}    & \begin{tabular}[c]{@{}l@{}}Pipe\\ Only\end{tabular}          & N/A                                                         & N/A                                                      \\ \hline
VMM                                                                   & \begin{tabular}[c]{@{}l@{}}Atlas \\ Muon Micromegas \\ Thin \\ gap\end{tabular} & BNL                                                                       & \begin{tabular}[c]{@{}l@{}} A-S-D-\\PA-TOT\end{tabular}                                                                                             & 64   & 25-200 ns                                                      & wide                                              & 1-9                                                    & \begin{tabular}[c]{@{}l@{}} 1000 e \\ at 10pF\end{tabular}    & N/A                                                          & N/A                                                         & N/A                                                      \\ \hline
VMM2                                                                  & \begin{tabular}[c]{@{}l@{}}Atlas \\ Muon Micromegas \\ Thin gap\end{tabular}    & BNL                                                                       & \begin{tabular}[c]{@{}l@{}}+6bit \\ peak \\ FADC\\ 10 bit\\ slowADC\end{tabular}                         & 64   & 25-200 ns                                                      & wide                                              & 1-9                                                    & \begin{tabular}[c]{@{}l@{}} 1000 e \\ at 10pF\end{tabular}    & N/A                                                          & N/A                                                         & N/A                                                      \\ \hline
TIMEPIX2                                                              & \begin{tabular}[c]{@{}l@{}}GEM\\ Micromegas\end{tabular}                        & CERN                                                                      &                                                                                                          &      &                                                                &                                                   &                                                        &                                                               &                                                              &                                                             &                                                          \\ \hline
\end{tabular}
\end{sidewaystable}

Other chips that have extensively been used both with thermal and fast neutron detectors are the CARIOCA-GEM digital and self-triggered chips~\cite{CARIOCA}.
These chips (each with 8 channels) were realised in order to equip the LHCB-GEMs and have recently found several new applications.
As presented during the workshop~\cite{Croci_talk}, these chips have a maximum count rate capability per channel of about 10~MHz and have been used coupled with a custom made LNF-INFN FPGA motherboard.
The system of CARIOCA+FPGA gives the possibility both to reconstruct the 2D map of a padded GEM detector and to perform TOF measurement either by exploiting the internal feature of movable gate of the FPGA.
In addition if the Logical Signal coming from the CARIOCA chip is properly adapted it can be interfaced to already existing DAQ system, such as the DAE of the ISIS facility~\cite{DAE}.
New chips called GEMINI are being developed by INFN (MIB and LNF sections) in order to substitute and improve the CARIOCA performances.
Another very interesting chip that was presented during the workshop was the AGET+ASAD electronics.
This chip was tested in combination with MicroMegas detectors and is an analog and auto-triggered chip featuring 64 channels, a sampling rate of 100~MHz and a dynamic range from 120~fC to 10~pC.
In addition through this chip it is possible to supply high voltage.
This chip was used both to get signals from grounded anode strips and from mesh strips put at high voltages.
These gave the possibility to perform time-of-flight measurements at neutron sources.

\section{Conclusions}

During this second workshop dedicated to neutron detection with MPGD, we noticed a sharp improvement with respect to the first edition both regarding detector manufacturing techniques and performances.

The construction at Link\"{o}ping University (Sweden) of a new coating machine for enriched boron carbide deposition over large areas allows to realise new geometrical configurations of the detectors in order to increase detection efficiency.

We strongly think that the next workshop on this topic must involve instrument scientists working at spallation neutron sources in order to better understand their requirements and to realise highly integrated detection systems with performances better than present instrumentation.

\end{document}